\documentclass{cjour}
\usepackage{epsfig}

\begin{document}


\authorrunninghead{Dorband, Hemsendorf, Merritt}
\titlerunninghead{Systolic N-body codes}



\title{Systolic and Hyper-Systolic Algorithms for the Gravitational
N-Body Problem, with an Application to Brownian Motion} 

\author{Ernst Nils Dorband, Marc Hemsendorf, and David Merritt}
\affil{Department of Physics and Astronomy, Rutgers University,
New Brunswick, NJ}

%



\def\fun#1#2{\lower3.6pt\vbox{\baselineskip0pt\lineskip.9pt
  \ialign{$\mathsurround=0pt#1\hfil##\hfil$\crcr#2\crcr\sim\crcr}}}
\def\lap{\mathrel{\mathpalette\fun <}}
\def\gap{\mathrel{\mathpalette\fun >}}
\def\dvp{\langle\Delta v_{\parallel}^2\rangle}
\def\mh{M_{\bullet}}
\def\msun{M_{\odot}}


\abstract{ 
A systolic algorithm rhythmically computes and passes data through 
a network of processors.
We investigate the performance of systolic 
algorithms for implementing the gravitational
$N$-body problem on distributed-memory computers.
Systolic algorithms minimize memory requirements
by distributing the particles between processors.
We show that the performance of systolic routines
can be greatly enhanced by the use of
non-blocking communication, which allows particle
coordinates to be communicated at the same time
that force calculations are being carried out.
The performance enhancement is particularly great when
block sizes are small, i.e. when only a small fraction
of the $N$ particles need their forces computed in 
each time step.
Hyper-systolic algorithms reduce the communication
complexity from $O(Np)$, with $p$ the processor
number, to $O(N\sqrt{p})$, at the expense of increased
memory demands. 
We describe a hyper-systolic algorithm that will work
with a block time step algorithm and analyze its performance.
As an example of an application requiring large $N$, 
we use the systolic algorithm to carry out direct-summation
simulations using $10^6$ particles of the Brownian motion 
of the supermassive black hole at the center of the
Milky Way galaxy.
We predict a 3D random velocity of $\sim 0.4$ km s$^{-1}$ for
the black hole.

}

\begin{article}

\section{Introduction}
\label{sec:intro}

Numerical algorithms for solving the gravitational $N$-body problem 
have evolved along one of two lines in recent years.
Direct-summation codes compute the complete set of $N^2$ interparticle 
forces at each time step; these codes were designed for systems
in which the finite-$N$ graininess of the potential is important,
and are limited by their $O(N^2)$ scaling to moderate ($N\lap 10^5$)
particle numbers.
The best-known examples are the {\tt NBODY} series of codes introduced by
Aarseth \cite{Aarseth:99a}.
These codes typically use high-order schemes for integration 
of particle trajectories and avoid the force singularities at small 
interparticle separations either by softening, 
or by regularization of the equations of motion \cite{KS:65}.
A second class of $N$-body algorithms replace the
direct summation of forces from distant particles by an approximation scheme.
Examples are tree codes \cite{Barnes:89}, which reduce the number
of direct force calculations by collecting particles in boxes, 
and algorithms which represent the large-scale potential via a truncated 
basis-set expansion (e.g. \cite{Allen:90}) or on a grid (e.g. \cite{Miller:68}).
These algorithms have a milder, $O(N\log N)$ scaling for the force
calculations and can handle much larger particle numbers although at
some expense in decreased accuracy \cite{Spurzem:99}.

The efficiency of both sorts of algorithm can be considerably
increased by the use of individual time steps for advancing
particle positions, since many
astrophysically interesting systems exhibit a ``core-halo''
structure characterized by different regions with widely disparate
force gradients.
An extreme example of a core-halo system is a galaxy 
containing a central black hole \cite{Merritt:99}.
The efficiency of individual time steps compared with a global
time step has rendered such schemes standard elements  
of direct-summation codes (e.g. \cite{Aarseth:99b}).

Here we focus on direct-summation algorithms as implemented on
multi-processor, distributed-memory machines.
Applications for such codes include simulation of globular star
clusters, galactic nuclei, or systems of planetesimals orbiting a 
star. In each of these cases, values of $N$ exceeding 
$10^5$ would be desirable and it is natural to investigate parallel
algorithms.
There are two basic ways of implementing a parallel force computation
for $O(N^2)$ problems.

1. {\bf Replicated data algorithm.} All of the particle information
is copied onto each processor at every time step.  Computing
node $i$, $1\le i\le p$, computes the forces exerted by the entire set of
$N$ particles on the subset of $n_i=N/p$ particles assigned to it.

2. {\bf Systolic algorithm.} At the start of each time step,
each computing node contains only $N/p$ particles. The sub-arrays
are shifted sequentially to the other nodes where the partial forces
are computed and stored. After $p-1$ such shifts, all of the force
pairs have been computed. 

\noindent
(The term ``systolic algorithm'' was coined by H. T. Kung  
\cite{Kung:78,Kung:82} by analogy with blood circulation.)
Both types of algorithm exhibit an $O(Np)$ scaling in communication
complexity and an $O(N^2)$ scaling in number of computations.
The advantage of a systolic algorithm is its more efficient use of
memory: since each processor stores only a fraction $1/p$ of the
particles, the memory requirements are minimized and a larger number
of particles can be integrated.
Other advantages of systolic algorithms include elimination of
global broadcasting, modular expansibility, and simple and regular
data flows \cite{Kung:82}.

The performance of a systolic algorithm suffers, however, whenever
the number of particles on which forces are being computed is less than
the number of computing nodes.
This is often the case in core-halo systems
since only a fraction of the particles are advanced 
during a typical time step.
As an extreme example, consider the use of a systolic algorithm 
to compute the total force on a {\it single} particle due to
$N$ other particles.
Only one processor is active at a given time and the total
computation time is
\begin{equation}
N\tau_f + p(\tau_l + \tau_c) 
\end{equation}
where $\tau_f$ is the time for one force calculation,
$\tau_l$ is the latency time required for two processors to
establish a connection, and $\tau_c$ is the interprocessor
communication time.
Thus the algorithm is essentially linear and no advantage is
gained from having multiple processors.

An efficient way to deal with the problem of small group
sizes in systolic algorithms is via {\it nonblocking communication},
a feature of MPI that allows communication to be put in the
background so that the computing nodes can send/receive data
and calculate at the same time \cite{MPI:98}.
In a nonblocking algorithm, the time per force loop for a single
particle becomes
\begin{equation}
{N\tau_f\over p} + p(\tau_l + \tau_c).
\end{equation} 
The second term is the waiting time for the last computing node
to receive the particle after $p$ shifts.
The first term is the time then required to compute the forces
from the subset of $N/p$ particles associated with the last node.
As long as the calculation time is not dominated by interprocessor
communication, the speedup is roughly a factor of $p$ compared with
the blocking algorithm.

Here we discuss the performance of systolic algorithms 
as applied to systems with small group sizes, 
i.e. systems in which the number of particles whose
positions are advanced
during a typical time step is a small fraction of the total.
Section 2 presents the block time step scheme and its implementation
as a systolic algorithm.
Section 3 discusses the factors which affect the algorithm's
performance, and Section 4 presents the results of performance
tests on multiprocessor machines of blocking and nonblocking algorithms.
Section 5 presents a preliminary discussion of ``hyper-systolic''
algorithms with block time steps,
which achieve an $O(N\sqrt{p})$ communication complexity
at the cost of increased memory requirements.
Finally, Section 6 describes an application of our systolic 
algorithm to a problem requiring the use of a large $N$:
the gravitational Brownian motion of a supermassive black hole
at the center of a galaxy.

\section{Algorithm}
\label{sec:algorithm}

In a direct-force code, the gravitational force acting on particle $i$
is
\begin{equation}
\mathbf{F}_i = m_i\mathbf{a}_i = -Gm_i\sum_{k=1}^N 
	{m_k \left(\mathbf{r}_i - \mathbf{r}_k\right) 
	\over |\mathbf{r}_i-\mathbf{r}_k|^3}
\label{eq:nbforce}
\end{equation}
where $m_i$ is the mass of the $i$th particle and $\mathbf{r}_i$ its
position; $G$ is the gravitational force constant. The summation excludes
$k=i$.

The integration of particle orbits is based on the fourth-order
Hermite scheme as described by Makino and Aarseth \cite{Makino:92}. 
We adopt their formula for computing the time-step of an individual 
particle $i$ at time $t_\mathrm{now}$,
\begin{equation}
\Delta t_{i} = \sqrt{ 
	\eta{|\mathbf{a}(t_\mathrm{now})| 
	|\mathbf{a}^{(2)}(t_\mathrm{now})| +
	|\mathbf{\dot a}(t_\mathrm{now})|^2
	\over |{\bf\dot a}(t_\mathrm{now})| 
	|\mathbf{a}^{(3)}(t_\mathrm{now})| + 
	|\mathbf{a}^{(2)}(t_\mathrm{now})|^2 }}.
\label{eq:ind-ts}
\end{equation}
Here $\mathbf{a}$ is the acceleration of the $i$th particle, 
the superscript $(j)$ denotes
the $j$'th time derivative, and $\eta$ is a dimensionless constant of
order $10^{-2}$; we typically set $\eta=0.02$. With a definition of
individual time-steps as in equation (\ref{eq:ind-ts}) the computing
time for the integration of typical gravitational systems is
significantly reduced in comparison with codes which do a full force 
calculation at each integration step. Since the particle positions 
$\mathbf{r}_k$
must be up-to-date in equation (\ref{eq:nbforce}) they are predicted
using a low order polynomial. This prediction takes less time if
groups of particles request a new force calculation at large
time intervals, rather than if single particles request
it in small time intervals. 
For this reason, an integer $n$ is chosen such that
\begin{equation}
\left({1\over 2}\right)^{n} \; \le \; \Delta t_{i}
\; < \; \left({1\over 2}\right)^{n-1}
\end{equation}
with $\Delta t_i$ given by equation (\ref{eq:ind-ts}).
The individual time step is replaced by a block time step 
$\Delta_b t_{i}$, where
\begin{equation}
\Delta_b t_{i} = \left({1\over 2}\right)^n.
\end{equation}

We implement the systolic force calculation in either blocking or
nonblocking mode.  For the nonblocking calculation, we have to add a
buffer for storing the incoming positions and masses, one for the
outgoing positions and masses, and one compute buffer. The maximal
size of these buffers is defined by the largest particle number on one
processing element.  We also need input, output and compute buffers
for the forces and the time-derivatives of the forces. Since data
synchronization is not critical with blocking communication, extra
input data buffering is not necessary. A positive side effect of the
buffering strategy is that data access is far more ordered than in
other implementations of the Hermite scheme. As a result, the number
of cache misses is reduced, optimizing performance.

\begin{figure}
\includegraphics[width=\textwidth]{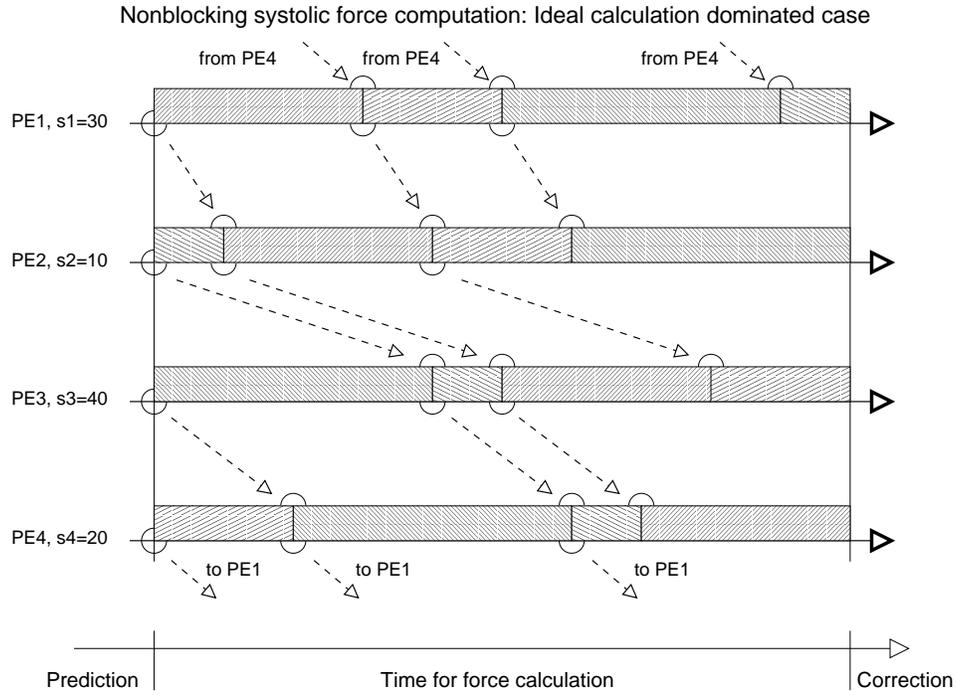}
\caption{Workflow diagram for an ideal, calculation-dominated
nonblocking systolic force computation. Time increases to the
right; the bold-faced arrows represent the work of each of
the computing nodes, assumed here to be four. The dashed arrows 
indicate the flow of the position information between the nodes. 
Circles indicate the points when each processor switches from computing 
the partial forces from one subgroup to the next subgroup, including
the time to finalize one communication thread and initialize the next one. 
Vertical lines represent barriers which can only be passed when all 
processes reach the same state of execution.}
\label{fig:nonbl-systlc-loop}
\end{figure}

We arrange all processors in a ring-like structure, so that each
processor has a right and a left neighbor. 
For the integrator to work, the individual block time-step $\Delta_b t_i$
and the time of the last integration $t_\mathrm{last}$ must be
defined. Either the initialization or the integration method are
required to compute these two quanities.

\begin{nonumalgorithm}[Find new group]
\textbf{Search} all particles $i$ for the smallest $t_{\mathrm{min},p}%
= \Delta_b t_{i,p} + t_{\mathrm{last},i,p}$  
\ on each computing node $p$.
\\ 
\textbf{Do a global reduce} so that each processor knows the global
\ minimum $t_\mathrm{min} = \min(t_{\mathrm{min},p})$. Set the simulated
\ time to $t_\mathrm{now} = t_\mathrm{min}$.
\\
\textbf{Find} the particles with $\Delta_b t_{i,p} +%
t_{\mathrm{last},i,p} = t_\mathrm{now}$ 
\ and store their index $i$ in a list. 
\\
\textbf{Predict} the positions and velocities at the time
\ $t_\mathrm{now}$ for the local particles on each node. 
\end{nonumalgorithm}

Each processor will select a subgroup $s_p$ of the block size
$s = \sum \; s_p$. The systolic shift with blocking communication 
is implemented as follows:

\begin{nonumalgorithm}[force loop (blocking communication)]
\textbf{Copy} the positions of the subgroups $s_p$ into the
\ compute buffer on each node.
\\
\textbf{foreach} j in $s_p$\label{alg:forceloop}
\ Do the force calculation with respect to all local particles. 
\textbf{end foreach}
\\
\textbf{Wait} for all processors to finish their work.
\\
\textbf{Copy} the masses, positions, and partial forces to the output
buffer. 
\\ 
\textbf{Send} the output buffer to the right neighbor and store the
\ data from the left neighbor in the compute buffer. 
\\ 
\textbf{Continue} at line \ref{alg:forceloop} unless $p$ shifts have
\ been done and the forces returned to their originating processor.
\end{nonumalgorithm}

Utilizing nonblocking communication, the systolic loop can gain
significant performance since it is possible to transfer data while
the force calculation is ongoing.
The partial forces follow one cycle behind the positions.

\begin{nonumalgorithm}[force loop (nonblocking communication)]
\textbf{Copy} the positions of the subgroups $s_p$ into the
\ compute buffer and into the output buffer on each node.
\\
\textbf{if} nloops between 1 and $p - 1$:\label{alg:posshift}
\ \textbf{Initiate} the data transfer for the positions and masses to
\   the right node and allow data to be sent from the left neighbor
\   to the input buffer.
\textbf{end if}
\\
\textbf{if} nloops between 2 and $p$:
\ \textbf{Initiate} the data transfer for the forces to the right node
\   and allow data to be sent from the left neighbor to the input buffer.
\textbf{end if}
\\
\textbf{foreach} j in $s_p$
\ Do the force calculation with respect to all local particles. 
\textbf{end foreach}
\\
\textbf{if} nloops between 1 and $p - 1$:
\ \textbf{Wait} for the data transfer of particles to finish.
\textbf{end if}
\\
\textbf{if} nloops between 2 and $p$:
\ \textbf{Wait} for the data transfer of the forces to finish.
\  Add the partial forces to the incoming forces.
\textbf{end if}
\\
 \textbf{Increment} nloops and continue at line \ref{alg:posshift}
\ unless $p$ shifts have been made. 
\\
\textbf{Shift} the forces to the right neighbor. 
\end{nonumalgorithm}

Figure \ref{fig:nonbl-systlc-loop} shows an idealized workgraph of the
nonblocking systolic force computation,
assumed to be completely calculation-dominated, which would be the
case for a computer system with zero latency and infinite
bandwidth. 
In the example shown in Figure \ref{fig:nonbl-systlc-loop},
processor one finds $s_1 = 30$, processor two $s_2 = 10$, processor three
$s_3 = 40$, and processor four $s_4 = 20$ particles due for the force
calculation. The thin vertical lines represent barriers which all
processes can only pass together. The circles represent points in
time at which a process changes from the computation of subgroup $s_i$ to
the next subgroup. Since all communication is in the background, 
switching between the subcalculations is very fast and the
processing elements do not idle. The dashed arrows represent the
data flow between processors; they begin at the sending point and
terminate where the reception is finalized. In Figure 
\ref{fig:nonbl-systlc-loop}, communication between
processing elements one and two must be carried out
in a very short time as indicated by the steep inclination of
the arrows. This is why, of all communications in the force loop, the
transfer of $s_3$ between these two processors requires the highest
bandwidth.
In the ideal case, however, all processes finalize the
force calculation at the same time as is shown by the second
vertical bar. We discuss in
the following section how closely we can reach this ideal in typical
applications. 

\section{Factors affecting the performance}
\label{sec:expectation}

\subsection{General performance aspects}

The performance of systolic force calculations is dependent on the
calculation speed of each processing element and on the bandwidth of
the inter-process communication network. Also important is the
parallelism of the calculation and the load imbalance. However,
predicting these latter two quantities requires precise knowledge of the
distribution of work on the nodes and of numerous machine parameters.
Some of these parameters might also be dependent on the overall usage
of the parallel computer. For these reasons we restrict our
estimates to an optimal and a worst-case scenario. 

Let $N$ be the total number of particles and $s$ the 
number of particles whose coordinates are to be advanced 
during the current integration step.
The particles to be integrated are distributed over $p$ subgroups of 
size $s_i$, $1 \leq i \leq p$,
such that the $i$th processor contributes $s_i$ particles.
Thus
\begin{eqnarray}
  \label{eq:s_sum}
  \sum_{i=1}^p s_i = s, \\
  \langle s_i \rangle = {s \over p}.
\end{eqnarray}
Note that the $s_i$ need not be equal, since the
number of particles due to be integrated
may be different on different processors.
We assume that the time spent on force calculations 
is a linear function of the group size $s_i$, and similarly that
the communication time is a linear function of the amount of 
transported data.

The time required for the force calculation can be estimated as follows.
Let $\tau_f$ be the time required to do one pairwise force calculation.
The total number of force calculations that one processor has to
do per full force loop is $(N/p)s$, since the processor calculates the
force of its $(N/p)$ own particles against the $s$ particles of the group.
The total time for one full force loop in the calculation-dominated 
regime is therefore
\begin{equation}
  \label{eq:t_f}
  t_f = {N s \over p} \tau_f.
\end{equation}

The time necessary for communicating a subset of particles 
$s_i$ from one processor
to its neighboring one can be estimated as follows. Let $\tau_c$ be
the time required to transfer the information for one particle to the
next processor. The latency time $\tau_l$ is the time it takes to set
up the communication between two processors. Within a single
integration step, each processor has to establish $p$ connections and
has to send $s$ particles. The total communication time is therefore:
\begin{equation}
  \label{eq:t_c}
  t_c = p \tau_l + s \tau_c.
\end{equation}

\subsection{Blocking communication}
\label{sec:comparison_nonblocking}

If the communication is not delegated to a communication
thread, the force calculation phase for each subgroup $s_i$ is
followed by a communication phase. 
We first consider the case that all group sizes $s_i$ are the same.
Then the total time is just the sum of $t_f$ and $t_c$, or
\begin{equation}
t={Ns\over p}\tau_f + p\tau_l + s\tau_c.
\end{equation}
The optimal processor number is obtained when $dt/dp=0$, or
\begin{equation}
p_{\mathrm{opt}} = \sqrt{{\tau_f\over\tau_l}Ns}
\end{equation}
and
\begin{equation}
t_{\mathrm{opt}} = s\tau_c + 2\sqrt{\tau_f\tau_lNs}.
\end{equation}

We now consider the case that the block sizes $s_i$ are not equal.
Each shift of particles must now wait for the processor with the
largest block size $s_{max}$ to complete its calculation
and communication. So for each shift, the time $t_s$ is
\begin{equation}
t_s = {N\over p}s_{max}\tau_f + \tau_l + s_{max}\tau_c
\end{equation}
and after $p$ shifts,
\begin{equation}
t = Ns_{max}\tau_f + p\tau_l + ps_{max}\tau_c.
\end{equation}
Defining
\begin{eqnarray}
s_{max} &=& \langle s_i\rangle + \Delta_{max}s_i \nonumber \\
&=& {s\over p} + \Delta_{max}s_i,
\end{eqnarray}
we can write:
\begin{eqnarray}
t = {Ns\over p}\tau_f + p\tau_l + s\tau_c + N\Delta_{max}s_i\tau_f +
	p\Delta_{max}s_i\tau_c.
\label{eq:tblocking}
\end{eqnarray}
Setting ${d t/d p} = 0$,
\begin{equation}
p_{\mathrm{opt}} = \sqrt{\frac{\tau_f N s} 
	{\tau_l + \tau_c \Delta_{max}s_i}},
\end{equation}
and the time is
\begin{equation}
t_{\mathrm{opt}} = 2 \sqrt{N s \tau_f (\tau_l + \tau_c
	\Delta_{max}s_i)} + s \tau_c + N \Delta_{max}s_i \tau_f.
\end{equation}

\subsection{Nonblocking communication}

On a system which allows concurrent communication and calculation, the
systolic algorithm can become more efficient, since the effective cost
of communication is reduced. In this case the communication routines
return immediately after initializing the communication process. 
As described in more detail in section
\ref{sec:algorithm}, each processor has to perform the following steps
$p$ times for each integration step:
\begin{enumerate}
\label{it:steps}
\item Do the force calculation for the $s_i$ particles of one subgroup. 
\item Simultaneously send the $s_i$ particles to the next processor.
\item Simultaneously receive the $s_{i-1}$ particles from the previous 
processor.
\end{enumerate}

Each of these steps starts at the same time, but they might
take different amounts of time to finish.
Since (2) and (3) behave in a quite
similar way, we treat them together and call them
\emph{communication}. Step (1) is called \emph{calculation}. A
system in which the calculation takes more time than the communication
is called a \emph{calculation-dominated system}, and a system in which
communication is dominant we call a \emph{communication-dominated
system}. Our goal is to derive approximate expressions for the total
time, calculation plus communication, per integration step and to
minimize this time.

\begin{figure}
\includegraphics[width=\textwidth]{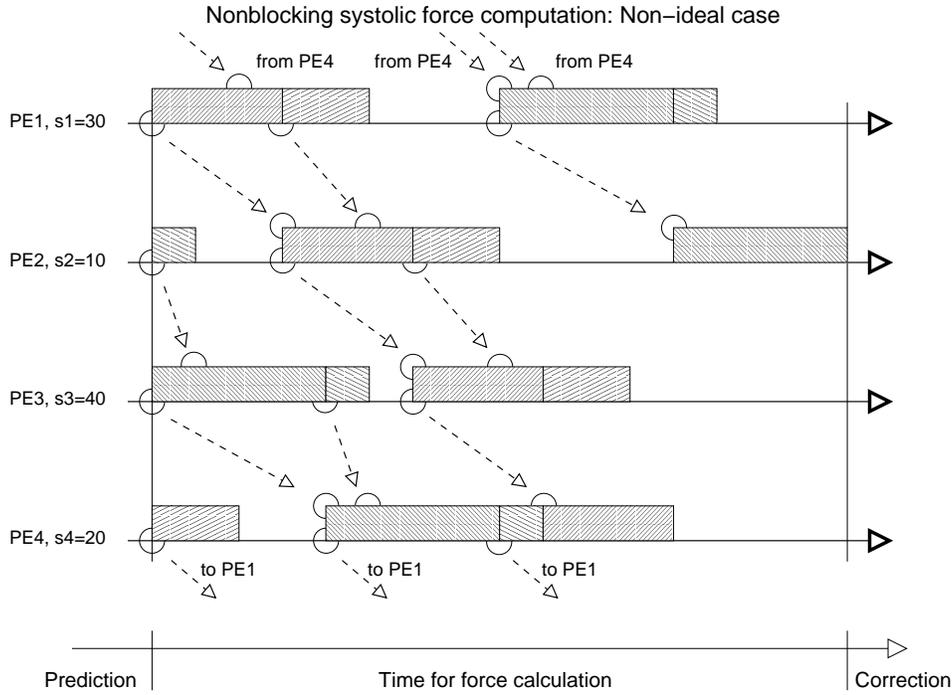}
\caption{Workflow diagram for a mixed communication- and 
calculation-dominated systolic force calculation. In this example 
we have chosen
the communication time to be a linear function of the calculation time
with $t_{c,i} = t_{f,i}$ and $\tau_l = 0$. The hatched blocks symbolize
times when the processors perform the force calculations. The circles
at the lower ends of these blocks denote times when the
communication of positional data is initiated. The ones at the upper
end denote the times when the communication has completed. The dashed
arrows show the dataflow. In order to simplify this graph, the
communication of partial forces is omitted.} 
\label{fig:nonbl-systlc-slow}
\end{figure}

Figure \ref{fig:nonbl-systlc-slow} shows a communication-dominated
system. In this example, $\tau_l = 0$ and $\tau_c =
\tau_f$. With this communication speed and the very uneven
distribution of subgroups $s_i$ on the processors, it is not possible
to ensure a continuous force calculation. The time to compute
the forces is dominated by the communication of the largest
group ($s_3$ in Figure \ref{fig:nonbl-systlc-slow}). With this
scenario we are able to define a threshold for the communication
speed, which ensures that processors are not waiting for the
communication to finish. Let $t_{c,max}$ be the processor-to-processor 
communication time for the largest subgroup $s_{max}$ and $t_{f,min}$ 
the force calculation time for the smallest subgroup $s_{min}$. 
Continuous calculation can be expected when
\begin{equation}
t_{c,max} \le  t_{f,min}.
\label{eq:threshold}
\end{equation}

The values for the latency time and the throughput are dependent on
the specific MPI implementation. Cray T3Es typically
have a two-mode implementation with algorithms for small
and for large messages. On the T3E-900 in Stuttgart, the sustained
average latency and throughput has been measured to be $\tau_l = 6
\;\mu s$ for the routines \verb'MPI_Send' and
\verb'MPI_Recv' \cite{Rabenseifner:99}. The MPI implementation on this
machine actually has a four-mode scheme which provides the following
measured bandwidths $\beta$:
\begin{eqnarray*}
\beta_{1} & \ge & 220 \, \mathrm{Mbyte/sec} \quad \mathrm{for} \quad
L_m \; \ge \; 8 \, \mathrm{kbyte}, \\
\beta_{2} & \ge & 300 \, \mathrm{Mbyte/sec} \quad \mathrm{for} \quad
L_m \; \ge \; 64 \, \mathrm{kbyte}, \\
\beta_{3} & \ge & 315 \, \mathrm{Mbyte/sec} \quad \mathrm{for} \quad
L_m \; \ge \; 256 \, \mathrm{kbyte}.
\end{eqnarray*}
\noindent
The quantity $L_m$ denotes the message length. These figures show that
the actual throughput is dependent on
the machine architecture, load and MPI implementation
\cite{Rabenseifner:99}. 

\begin{figure}
\includegraphics[width=\linewidth]{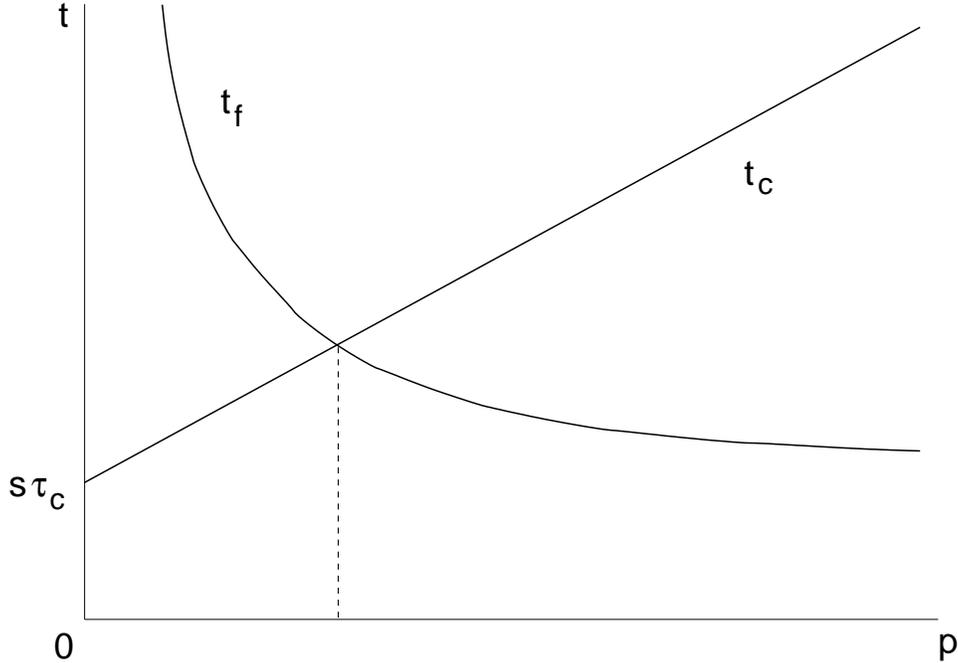}
\caption{Communication and calculation time as a function of
processor number $p$ for fixed $N$ as derived from equations (\ref{eq:t_f})
and (\ref{eq:t_c}). This figure is only valid under the assumption of 
equal $s_i$.} 
\label{fig:popt}
\end{figure}

Assuming all $s_i$ to be equal, a rough estimate for an optimal
processor number can be given. 
Figure \ref{fig:popt} shows that there is an optimum value
$p=p_{\mathrm{opt}}$ which minimizes the total time in this case. We
find $p_{\mathrm{opt}}$ by setting (\ref{eq:t_f}) equal to
(\ref{eq:t_c}):
\begin{equation}
  p_{\mathrm{opt}} = {-s \tau_c + \sqrt{s^2 \tau_c^2
  + 4 \tau_l \tau_f N s} \over 2 \tau_l}
  \label{eq:p_opt} 
\end{equation}
and the total time is
\begin{equation}
  t_{\mathrm{opt}} = {s\tau_c\over 2} + 
	{1\over 2}\sqrt{s^2\tau_c^2+4\tau_l\tau_fNs}.
 \label{eq:t_opt}
\end{equation}
Equations \ref{eq:t_opt} and \ref{eq:p_opt} become far simpler on
machines having zero latency. With $\tau_l = 0$ we find
\begin{eqnarray}
  p_{\mathrm{opt}} & = & N \, \frac{\tau_f}{\tau_c},\\
  t_{\mathrm{opt}} & = & s \, \tau_c.
\end{eqnarray}
This means that the systolic force loop parallelizes extremely well on such
an idealized computer so that the computing time is only dominated by
the performance of the intercommunication network. However this
equation also allows the use of more processors than particles if the
communication is faster than the force calculation. We
give a more detailed picture of the situation in the following. 

In $N$-body systems with a low central concentration, 
i.e. small core density, $s$ is usually proportional to 
$N^{2/3}$ \cite{Makino:92}. In our benchmarks discussed below, 
however, we find power-law indices of $0.55 \pm 0.05$ for the 
Plummer model, $0.437 \pm 0.105$ for the Dehnen model, and 
$0.492 \pm 0.110$ for the Dehnen model with a black hole. 
For this reason, we estimate the
groupsizes to be $s \approx \xi N^{1/2}$. The constant $\xi$ can be derived
for each type of dataset from the measured values for $s$ in Table
\ref{tab:groupsizes}. 

In our scheme,  a given processor may be calculation- or 
communication-dominated depending on the value of $s_i$ that 
is currently assigned to it.  
We can compute the time per integration loop in this more
general case by again focussing on just one processor, which carries
out $p$ operations of force calculation and communication per loop.
Each of these $p$ operations will be either calculation- or
communication-dominated, requiring either a time of $t_{f,i}$ or
$t_{c,i}$. With $1 \; \leq \; i  \; \leq \; p$,
\begin{eqnarray}
  \label{eq:t_i_f_c}
  t_{f, i} & = & {N \over p} s_i \tau_f, \\
  t_{c, i} & = & \tau_l + s_i \tau_c.
\end{eqnarray}

With the notion of the threshold in equation (\ref{eq:threshold}),
which guarantees a continuous calculation phase, we have a completely
calculation-dominated system. Thus
\begin{eqnarray}
  t_\mathrm{calc} & = & \sum\limits_{i=1}^p \, t_{f,i}, \\
		  & = & \frac{N}{p} \, s \tau_f.
  \label{eq:scl-opt}
\end{eqnarray}
It is not trivial, though, to define an optimal processor number,
since the distribution of the $s_i$ and the size of $s$ is a dynamical
quantity of the integrated particle set. The result from equation
(\ref{eq:scl-opt}) describes the ideal scaling for the force
calculation which we use below as a basis of comparison with the
benchmark data.

If the threshold (\ref{eq:threshold}) is not fulfilled, a significant
share of the computation time $t$ comes from the communication. Assuming
a worst-case scenario, where all particles of the group $s$ are found
on only one processor, the computation time becomes maximal:
\begin{eqnarray}
  t_\mathrm{comm} & = & t_{f,max} + \sum\limits_{i=1}^{p} \,
	  t_{c,max}, \\
                  & = & t_{f,max} + p t_{c,max}.
  \label{eq:scl-bad}
\end{eqnarray}
The overall calculation time for the situation shown in Figure
\ref{fig:nonbl-systlc-slow} follows equation (\ref{eq:scl-bad}). 
We use this equation below to estimate the minimal scaling
behavior expected from our code.

\subsection{Comparing the performance of the two methods}
\label{sec:the_effect_of_small_group_sizes}

In comparison with the nonblocking algorithm, the blocking scheme does
not fall behind in terms of the parallelism. However, as long as the
force computations remain calculation-dominated in the
nonblocking scheme, minor variations between the $s_i$ can be levelled
out so that perfect load balancing is guaranteed. The blocking scheme
is not flexible in this regard so that it builds up a penalty of the
order $n \Delta_\mathrm{max} s_i \tau_f$ per force loop. Since
equation (\ref{eq:tblocking}) shows that there are a few particles in
the blocking scheme which are treated with a serial performance, the
overall parallel efficiency is reduced. This means that a systolic loop
will only give satisfying performance when load balancing is guaranteed
either on a per processor level by keeping all $s_i$ the same, or by
applying nonblocking communication. This is why a nonblocking
communication scheme is superior to the blocking one.

Depending on the type of system integrated, the number of
available processors, the performance of the computer, and the
total number of particles, the work load for the processors might
become very small. Small group sizes are less likely to be distributed
evenly on the processes; we expect an approximately Poisson distribution, or
\begin{equation}
  \Delta_{max}s_i \approx \sqrt{s_i}.
  \label{eq:poisson}
\end{equation}

For the systolic algorithm, large problem sizes, together with
group sizes that are larger in the mean than the number of
available processors, ensure efficient parallel performance with a systolic
algorithm. In this case one can expect an ideal linear scaling of the
performance with processor number. For small problem sizes, 
the amount of communication governs the the performance. 

\subsection{More elaborate codes}

Parallelizing an individual block time step scheme is not trivial, as
the discussion above has shown. However, is has also been shown that
a fairly simple scheme can profit enormously from threaded, and
therefore nonblocking, communication. Since the overall communication
time is a linear function of the processor number in our simple
scheme, attempts have been made to reduce the number of
communications. The hypersystolic codes proposed by Lippert et
al.~\cite{Lippert:98,Lippert:98-b} and the broadcast method proposed
by Makino \cite{Makino:2001} can perform the force calculation by
using only $\sqrt{p}$ communication events. However, both methods
require identical group sizes on each computing node in order to
have efficient load balancing. 

Individual block timestep schemes select their particles to be
integrated under physical criteria defined by the simulated
system. This is why a load imbalance is unavoidable in general. 
Assuming there is no extra algorithm that provides a
perfect distribution of the group, the expected imbalance time would
be dependent on the particle distribution statistics in equation
(\ref{eq:poisson}). Thus
\begin{equation}
t_\mathrm{imb} = \frac{n \Delta_\mathrm{max} s_i}{\sqrt{p}} \tau_f
\end{equation}
\noindent
This implies that the nonblocking systolic code will outperform
nonbalanced hypersystolic or broadcast methods unless the
processor number is very large. 

\section{Measured performance}
\label{sec:performance}

We evaluated the performance of the algorithms in realistic
applications by using them to evolve particle models of spherical stellar
systems.  The evolution was carried out for approximately one crossing
time, and the benchmarks were based on timings for $2000$ integration
steps. The Hermite integration scheme adjusts the group sizes
automatically as described above. All experiments 
were carried out on the Cray T3E at the Goddard Space Flight Center. 
We compiled the executable from our C sources using the Cray standard C
compiler version 6.2.0.0 with no explicit optimization. 

\subsection{Initial conditions}
We consider three models representing spherical stellar systems
in equilibrium.
The Plummer model \cite{Plummer:11} has mass density and gravitational
potential
\begin{eqnarray}
\rho(r) & = & 
  \frac{3 G M}{4 \pi} \; \frac{b^2}{\left(r^2 + b^2\right)^{5/2}},\\
\Phi(r) & = & - \frac{G M}{\sqrt{r^2 + b^2}}.
\end{eqnarray}
Here $M$ is the total mass and $b$ is a scale length.
The many analytic properties of this model make it a common
test case for benchmarking.
For the present application, the most important feature of the
Plummer model is its low degree of central concentration and
its finite central density, similar to the density profiles
of globular star clusters.
The Dehnen family of models \cite{Dehnen:93} are characterized by a parameter
$\gamma$ that measures the degree of central concentration.
The density profile is
\begin{equation}
\rho(r) = \frac{(3 - \gamma) M}{4 \pi} 
   \; \frac{a}{r^{\gamma} \left(r + a\right)^{4-\gamma}}
\label{eq:dehnen}
\end{equation}
with $M$ the total mass as before and $a$ the scale length.
We chose a centrally condensed Dehnen model with $\gamma=2$, 
yielding a density profile
similar to those of elliptical galaxies with dense stellar nuclei.
The gravitational potential for $\gamma=2$ is
\begin{equation}
\Phi(r) = -\frac{G M}{a} \; \ln\left(\frac{r}{r + a}\right).
\end{equation}
The central density diverges as $r^{-2}$ and the gravitational force
diverges as $r^{-1}$.
Our third model had a density equal to that of (\ref{eq:dehnen}) with
$\gamma=2$ and an additional mass component 
consisting of a point particle at the center representing a supermassive 
black hole.
The mass $\mh$ of the ``black hole'' was $1\%$ of the stellar mass 
of the model.
This is similar to the black-hole-to-galaxy mass ratios observed in 
real galaxies \cite{Merritt:01}.

The initial particle velocities for the $N$-body realizations of
our models were selected from the unique, isotropic velocity distribution
functions that reproduce $\rho(r)$ in the corresponding potentials $\Phi(r)$.
For all of the models we adopted units in which the gravitational constant $G$
and the total mass of the the system $M$ were set to unity.
The scale length $a$ of the Dehnen model was also set to unity,
while the Plummer radius $b$ was chosen to be $3 \pi/ 16$. 
We computed two realizations for each particle number $N$. 
The total number of stars was $N=(16384, 32768, 65536, 131072)$. 

%
%

\begin{table}
\begin{tabular}{lrrrr}
Model & 131072 & 65536 & 32768 & 16384 \\ \hline
Plummer                      & 172 &128 &89 &54 \\
Dehnen, $M_\bullet = 0$  & 13.1 & 11.2 & 9.2 & 5.1 \\
Dehnen, $M_\bullet = 0.01$ &  6.7 &  4.1 & 2.6 & 2.5
\end{tabular}
\caption{Mean Group Sizes for the Benchmarks}
\label{tab:groupsizes}
\end{table}

After the integration over 2000 steps we computed the average group
size during the benchmark. Table \ref{tab:groupsizes} summarizes the
results. The centrally condensed Dehnen model systems require the use of far
smaller groups than the Plummer model.
This poses a severe challenge to systolic algorithms. 

\subsection{Performance of the force calculation}
Since the complexity of the force calculation is $O(N^2)$, 
compared to $O(N)$ for the integrator, it is critical to measure the impact of
parallelism on this task. We present comparisons of blocking and nonblocking
systolic codes on the three datasets described above. We count 70 floating 
point operations per pairwise force calculation in our implementation.

The particle blocks may have very different sizes, and accordingly the
efficiency will vary from force step to force step. For this reason we
measure performance by summing the time required to carry out all
force computations in a time interval corresponding to $2000$ steps,
then averaging.

\begin{figure}
\includegraphics[width=\linewidth]{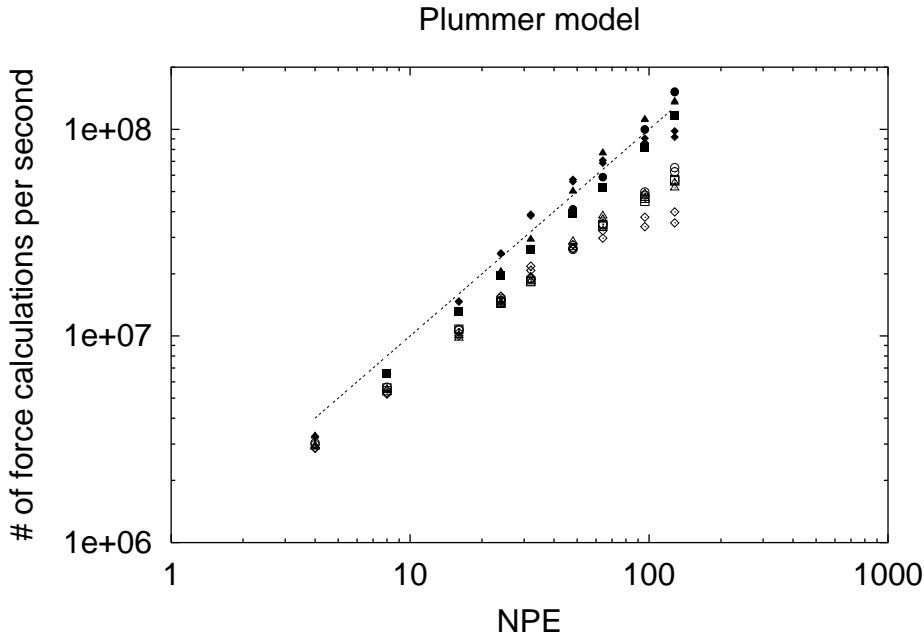}
\caption{Results of the benchmarks for the Plummer model initial
conditions. The number of pairwise force calculations per second 
has been plotted as a function of processor number $NPE$. 
Non-blocking and blocking algorithms are indicated via filled
and open symbols respectively, for the four different particle numbers
$N$ listed in Table 1.
The dashed line shows the expected performance in the case
of an ideal, calculation-dominated system (linear speedup).}
\label{fig:plm-bench}
\end{figure}

\begin{figure}
\includegraphics[width=\linewidth]{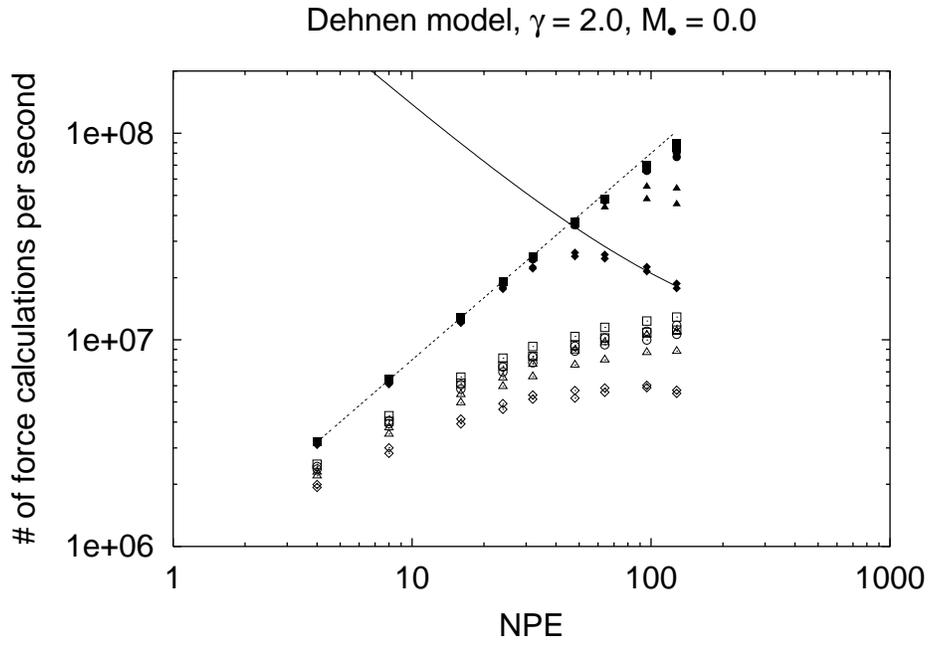}
\caption{Same as Figure \ref{fig:plm-bench}, for the Dehnen model
initial conditions. The dashed and solid lines show the expected scaling
for calculation- and communication-dominated systems.
The performance of the non-blocking algorithm is reduced for
small $N$ due to the small group sizes.
}
\label{fig:deh-bench}
\end{figure}

\begin{figure}
\includegraphics[width=\linewidth]{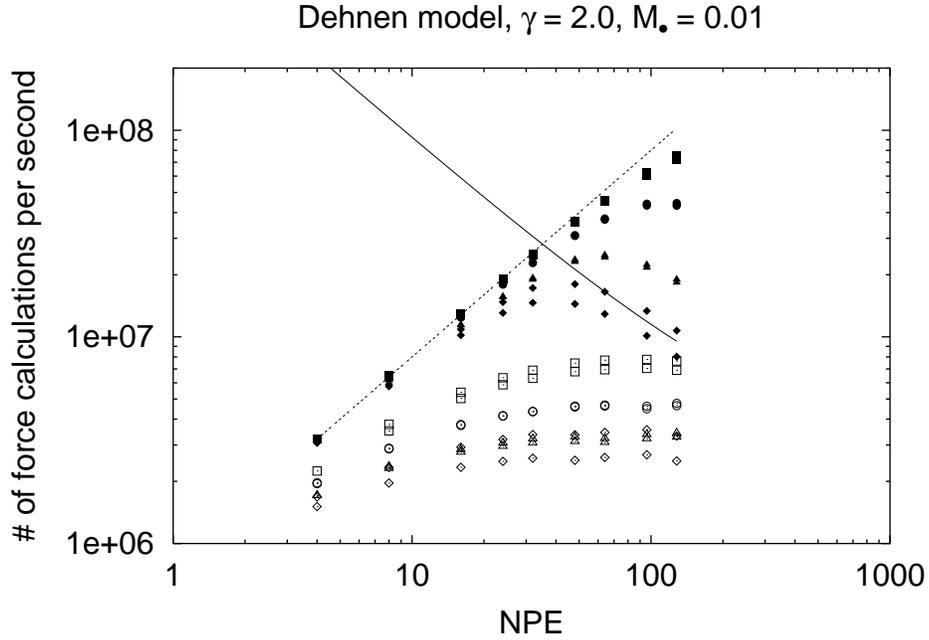}
\caption{Same as Figure \ref{fig:deh-bench}, for the Dehnen model
with a central point particle (``black hole'').
The influence of small group sizes is even more striking.
}
\label{fig:blk-bench}
\end{figure}

Figures \ref{fig:plm-bench}-\ref{fig:blk-bench} show the measured 
speedups, expressed in terms of the number of force calculations per second. 
In the case of the nonblocking algorithm, the expectation from 
equation (\ref{eq:scl-opt}), assuming a calculation-dominated system, 
is
\begin{equation}
t_\mathrm{calc} \propto \frac{1}{p},
\end{equation}
corresponding to a linear trend in Figures 
\ref{fig:plm-bench}-\ref{fig:blk-bench}.
This dependence is recovered for each of the three models when $N$
is large.
In the Dehnen-model runs with small $N$, the force calculation
becomes communication-dominated due to the small group sizes (Table 1)
and the performance is described by 
equation (\ref{eq:scl-bad}):
\begin{equation}
t_\mathrm{comm} \propto p
\end{equation}
corresponding to a $1/p$ dependence of the speedup in Figures 
\ref{fig:plm-bench}-\ref{fig:blk-bench}.
This effect is less pronounced for the Plummer models since the mean
group size is larger (Table 1).
The effect is strongest for the Dehnen models with
``black hole'' since they have the smallest group sizes
(Table 1).

In comparison with the runs involving Dehnen models, the
Plummer model runs show a peculiar behavior. For very low processor
number, the performance increases linearly. However for an intermediate
number of computing nodes, the performance shows a {\it super}-linear
gain until the curve becomes shallow for large processor numbers. This
super-linearity is a result of the way in which message passing is
implemented on the T3E, which switches to a different protocol for
small messages. The mean message size is inversely proportional to the
total number of processes so that we can profit from this change.  As
Table \ref{tab:groupsizes} shows, the mean group size (which is
proportional to the mean message size) is significantly smaller for
the Dehnen models explaining why the super-linearity does not occur.

The open symbols in Figures \ref{fig:plm-bench}-\ref{fig:blk-bench} 
show the results for the blocking systolic algorithm. In the Plummer model 
runs, the blocking code managed to calculate a little more than
half the number of force pairs per second compared with the nonblocking
code. The performance difference is much more dramatic in the Dehnen
model runs due to the small group sizes: the blocking code can
be as much as a factor of ten slower than the non-blocking code. 
The computing time is asymptotically constant in the Dehnen model runs
since the force calculation becomes increasingly serial for
small $s$, as given by equation (\ref{eq:tblocking}). For very large
$N$ we expect the $1/p$ scaling to take over. This does not appear in
the plots because of the inefficiency of the blocking force
calculation for small group sizes. 

\subsection{Performance of the integration}
The performance of the $N$-body code depends also on the efficiency
of the algorithm that advances the particle positions, which we call
the ``integrator.''
Here we give benchmarks for the full code including the integrator.
Our performance goal is to reach a linear increase of computer 
wall-clock time as a function of problem size for a constant simulation 
interval and optimal processor number. 
The unit of time in the simulations is fixed by the gravitational
constant (which is set to $G = 1$), the total mass, and the
adopted length scale via $[T]=(GM/R^3)^{-1/2}$ \cite{Heggie:86}. 
Our performance results are based on an integration using 2000 force
loops. The integrated time in model units is summarized in Table
\ref{tab:ttot-bench}.
We carried out two integrations for each model based on different
random realizations of the initial conditions.

\begin{table}
\begin{tabular}{l|cc}
Dataset & 1 & 2 \\ \hline 
Plummer & & \\
16384  & $4.57 \cdot 10^{-2}$ & $5.40 \cdot 10^{-2}$ \\
32768  & $3.35 \cdot 10^{-2}$ & $3.08 \cdot 10^{-2}$ \\
65536  & $1.90 \cdot 10^{-2}$ & $2.09 \cdot 10^{-2}$ \\
131072 & $1.08 \cdot 10^{-2}$ & $9.95 \cdot 10^{-3}$ \\
$\gamma = 2.0, M_\bullet = 0.0$ & & \\
16385  & $1.75 \cdot 10^{-3}$ & $1.59 \cdot 10^{-3}$ \\
32769  & $1.08 \cdot 10^{-3}$ & $1.35 \cdot 10^{-3}$ \\
65537  & $5.35 \cdot 10^{-4}$ & $6.68 \cdot 10^{-4}$ \\
131073 & $4.11 \cdot 10^{-4}$ & $3.26 \cdot 10^{-4}$ \\
$\gamma = 2.0, M_\bullet = 0.01$ & & \\
16385  & $3.99 \cdot 10^{-4}$ & $2.63 \cdot 10^{-4}$ \\
32769  & $1.53 \cdot 10^{-4}$ & $1.49 \cdot 10^{-4}$ \\
65537  & $1.30 \cdot 10^{-4}$ & $1.36 \cdot 10^{-4}$ \\
131073 & $4.52 \cdot 10^{-5}$ & $8.51 \cdot 10^{-5}$ \\
\end{tabular}
\caption{Integrated time after 2000 force loops for the datasets
used in the benchmarks}
\label{tab:ttot-bench}
\end{table}

\begin{figure}
\includegraphics[width=\linewidth]{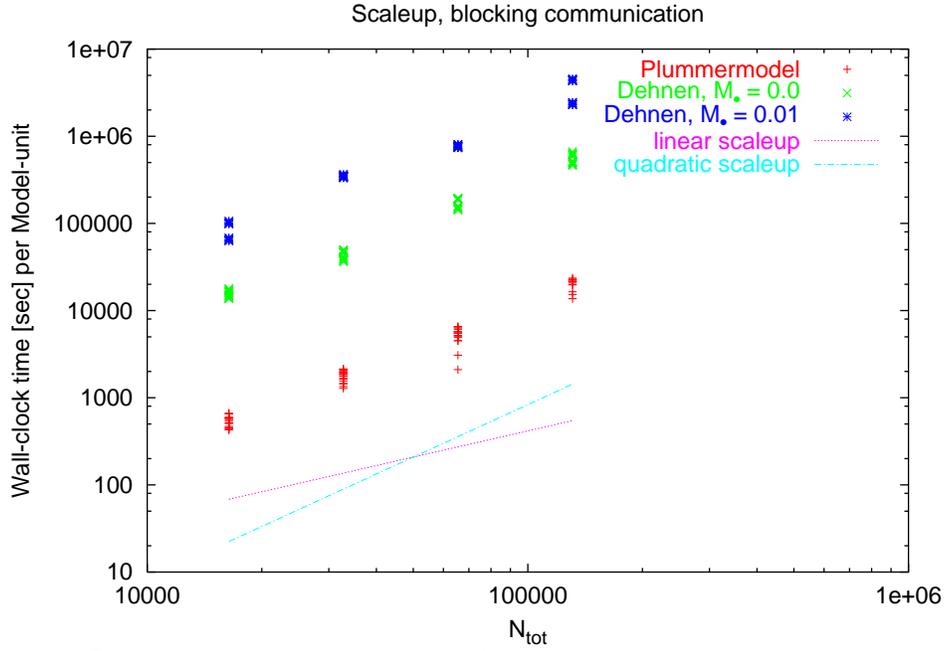}
\caption{Wall clock time in seconds per integrated model unit as a
function of particle number $N$ for the blocking $N$-body code. 
Different points at given $N$ show the performance for the various
values of NPE of Figures 4-6.
For comparison we plot lines following a linear and a
quadratic $N$-dependence. The lines are positioned in such a way 
that they mark the maximal speed of the nonblocking code.}
\label{fig:block-sclup}
\end{figure}

\begin{figure}
\includegraphics[width=\linewidth]{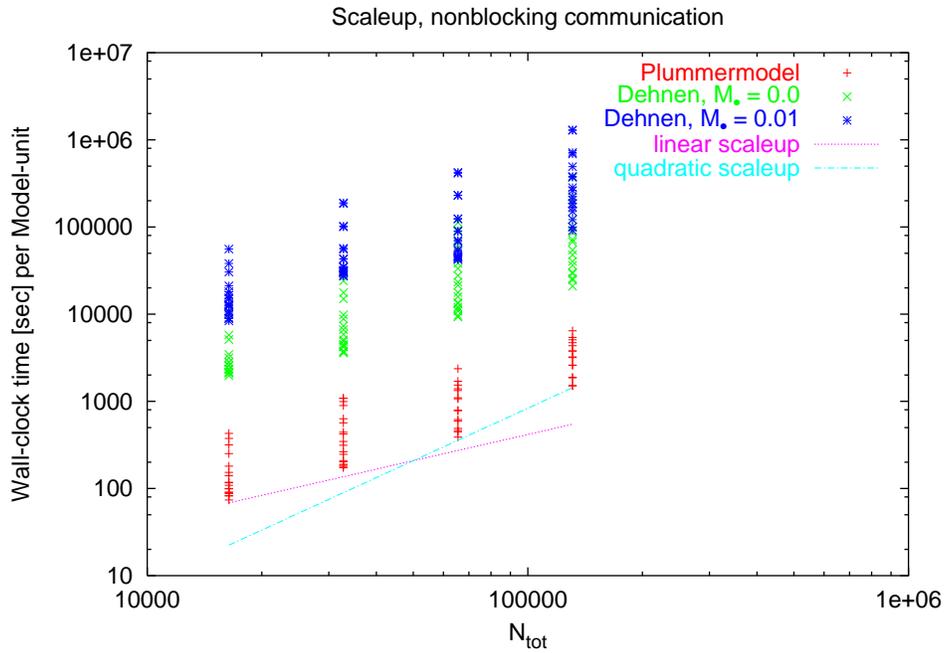}
\caption{Same as Figure \ref{fig:block-sclup} for the code with
nonblocking communication.}
\label{fig:nonbl-sclup}
\end{figure}

The performance of the blocking scheme is shown in Figure
\ref{fig:block-sclup}, for the same values of $N$ and NPE shown in
Figures \ref{fig:plm-bench}-\ref{fig:blk-bench}.  We now plot
performance as a function of $N$; runs with the same $N$ but different
NPE are plotted with the same symbol and their vertical scatter is an
indication of how well the code benefits from parallelism.  In the
blocking scheme, the points are very localized vertically revealing
that the code can not profit much from large processor numbers.
Larger problem sizes, however, are able to optimize the computations
in such a way that the scale-up is better than quadratic, the
scaling in traditional direct-force codes.
With the individual time step scheme described here we
expect a behavior of $N s$. Since we observe an $N^{1/2}$ behavior
for the mean group size, we can expect a scale-up following a $N^{3/2}$
power law in our benchmarks. We compare the maximal performance with
increasing work load in figures \ref{fig:block-sclup} and
\ref{fig:nonbl-sclup} for our code which shows a scale up between
$O(N^2)$ and $O(N^{3/2})$ on a serial machine. In our benchmarks we
consider any scaling better than that as indicating a benefit from parallel
computing. 

\begin{table}
\begin{tabular}{l|rr|rr}
Dataset & blocking, $b$ & $\Delta b$ & nonblocking, $b$ & $\Delta b$
\\ \hline 
Plummer & 1.65 & 0.03 & \textit{1.174} & \textit{0.112} \\
$\gamma = 2.0, M_\bullet = 0.0$ & 1.702 & 0.006 & 1.16 & 0.03 \\
$\gamma = 2.0, M_\bullet = 0.01$ & 1.612 & 0.003 & 0.9 & 0.1
\end{tabular}
\caption{Power-law fits of the two best performing runs in each
benchmark for the blocking and nonblocking communication runs.}
\label{tab:scalefits}
\end{table}

In order to quantify the scaleup, we fit the two, best-performing runs 
for each data set. 
We fit for $a$ and $b$ in the following function:
\begin{equation}
\frac{t_\mathrm{wc}}{T} = \frac{N_\mathrm{tot}^b}{a}
\end{equation}
The quantity $T$ the integrated time in model units and
$t_\mathrm{wc}$ is the wall clock time for the run.
Table \ref{tab:scalefits} summarizes the results for $b$.
Since the benchmarks using the Plummer model with the
nonblocking code did not reach their maximum speedup for $N=131072$ 
and $65536$, we estimated the scaleup from the runs with smaller
particle sets only. These values are italicized in Table
\ref{tab:scalefits}. 

When measuring the overall performance of the integrator, the
nonblocking communication scheme makes much better use of the T3E than
the blocking scheme:
the broad vertical bands in Figure \ref{fig:nonbl-sclup} show that the
integration speed profits from increasing processor number. The two
lines representing a linear and a quadratic behavior are placed at the
optimal speed levels for the runs integrating a Plummer model. For all
three types of model, a nearly linear scaleup can be observed. The
exceptions are the datasets representing a Plummer model with 131072
and 65536 particles. The reason for this is no bottleneck. In fact,
the nonblocking code scales so well that we could not reach the optimal
performance with our maximal number of 128 processing elements in the
benchmarks.

\subsection{Nonblocking vs. Blocking Communication}

While the implementation of a systolic force calculation is simpler
and more memory-efficient using blocking communication, 
we have found significant performance gains using a
nonblocking algorithm. The results of our force calculation benchmarks,
displayed in Figures \ref{fig:plm-bench}-\ref{fig:blk-bench}, show up
to ten times better performance for a code implementing a nonblocking
systolic force calculation over one which applies a blocking
scheme. The gain is greatest when the typical particle group size is 
smallest. The integration as a whole can profit more from the
parallelization in the nonblocking scheme than in the blocking: we
observe a scaleup close to $N^{3/2}$ for the blocking scheme, while
the nonblocking scheme clearly reaches a linear scaling. This means
that the nonblocking scheme is actually able to reduce the complexity
of the direct force integration to $O(N)$.

\section{Hypersystolic force calculation with individual block
time steps} 
\label{sec:hypersyst}
\subsection{Systolic and hyper-systolic matrices}
\label{sec:hyper-systolic-matrix}

Introduced by Lippert et al. \cite{Lippert:98,Lippert:98-b}, the
hyper-systolic algorithm reduces communication costs in $O(N^2)$
problems by improving the communication complexity from $O(Np)$
(systolic algorithm) to $O(N \sqrt{p})$. The price paid is an
increased need for memory, of order $O(\sqrt p )$
\cite{Lippert:98-b}.

The standard systolic algorithm is performed on a one-dimensional ring
of $p$ processors. In order to transfer all positions and forces,
$p$ shifts of all $N$ particles are required for each force
calculation. Thus the communication complexity is $O(N p)$.

By analogy with the systolic case, we define the vector of input data
$\mathbf{x} \equiv (x_1, x_2,... , x_p)$. In this definition the $x_i$ are
all data initially stored on processor $p_i$. These data are copied to a 
second vector $\mathbf{\tilde x}$ that is shifted between the processors. 
(Note our assumption here that the blocksize is $N/p$, i.e. that every 
particle needs its forces updated. We relax this assumption below.)
To calculate the total forces one has to take
into account all possible pairs of $x_i$ with $\tilde x_i$. 
To visualize the systolic algorithm, we write down a matrix that shows the
intermediate states of $\mathbf{\tilde x}$ at each step of the force
calculation. This matrix is called the ``systolic matrix'' $S$. The first
line shows $\mathbf{\tilde x}$ at shift zero, the $i$th line shows
$\mathbf{\tilde x}$ at shift $i$:
\begin{equation}
  \label{eq:systolic_matrix}
  S = \left ( 
    \begin{array}{c c c c c c c c}
      1 & 2 & 3 & 4 & 5 & 6 & 7 & 8 \\
      8 & 1 & 2 & 3 & 4 & 5 & 6 & 7 \\
      7 & 8 & 1 & 2 & 3 & 4 & 5 & 6 \\
      6 & 7 & 8 & 1 & 2 & 3 & 4 & 5 \\
      5 & 6 & 7 & 8 & 1 & 2 & 3 & 4 \\
      4 & 5 & 6 & 7 & 8 & 1 & 2 & 3 \\
      3 & 4 & 5 & 6 & 7 & 8 & 1 & 2 \\
      2 & 3 & 4 & 5 & 6 & 7 & 8 & 1 \\
    \end{array} \right ).
\end{equation}
Local computations are done within one column of this matrix. One
notices that there is some redundancy of vertical pairs: for instance,
the pair $(1, 2)$ occurs $8$ times. As a
result of this redundancy it is not necessary to perform all of the
shifts in order to get every possible pair of elements of $\mathbf{x}$. 
In fact it would be sufficient to perform just three shifts, so that we get a
matrix with just four rows. This matrix is called the ``hyper-systolic
matrix'' $H$ and is
\begin{equation}
  \label{eq:hyper-systolic_matrix}
    H = \left ( 
    \begin{array}{c c c c c c c c}
      1 & 2 & 3 & 4 & 5 & 6 & 7 & 8 \\
      8 & 1 & 2 & 3 & 4 & 5 & 6 & 7 \\
      7 & 8 & 1 & 2 & 3 & 4 & 5 & 6 \\
      5 & 6 & 7 & 8 & 1 & 2 & 3 & 4 \\
    \end{array} \right ).
\end{equation}
In contrast to the matrix $S$, which allows us to get all pairs 
by comparing the first row with one of the other rows,
not all of the pairs in $H$ have one member in the first row;
for example, the pair $(x_1, x_4)$ requires comparing rows $2$ and
$4$.
Therefore all of the shifted data in the hyper-systolic algorithm must
be stored on each node in order to compute the forces. This is the reason 
for the increased memory requirements in comparison with the systolic 
algorithm.

The hyper-systolic matrix $H$ can be characterized by a vector called
the hyper-systolic base $A_k = (0, a_1, ..., a_k)$ where $a_i$ gives the
stride of the $i$th shift and $k$ is the number of shifts that have to
be done. In our example of $H$ for $p=8$, $A_3$ would be:
\begin{equation}
  \label{eq:a_3}
  A_3 = (0, 1, 1, 2).
\end{equation}
To minimize the communication costs, one wants to minimize $k$, 
the number of shifts. This is a nontrivial problem and optimization 
techniques have been described for achieving this aim \cite{Lippert:97}.

In the following we show that the complexity of communication is
$O(n \sqrt{p})$ \cite{Lippert:98}. 
The minimum number of pairings required for the total force calculation 
is $p(p-1)/2$. The number of possible combinations within a column of 
$k+1$ elements is ${k+1 \choose 2}$, thus:
\begin{equation}
  \label{eq:proove1}
  {p(p-1) \over 2} \leq {k+1 \choose 2}  p = {(k+1)k \over 2}p.
\end{equation}
The solution for this inequality for $k \geq 0$ is
\begin{equation}
  \label{eq:solution}
  k \geq \sqrt{p-{3 \over 4}}-{1 \over 2},
\end{equation}
which we wanted to prove. However, since we do
$O(\sqrt{p})$ shifts and we have to save the intermediate data, we
need $O(\sqrt{p})$ more memory on each processor than with the
standard systolic algorithm.

\subsection{Hyper-systolic matrix with block time steps}
\label{sec:hyper-systolic-block}

The hyper-systolic algorithm described by Lippert et al. 
\cite{Lippert:98,Lippert:98-b} assumes that all $N(N-1)/2$ force
pairs are to be computed at each time step.  When dealing with block
time steps, however, only a subset of the full $N$ particles are
shifted at each time step, and a minimal basis like that of equation
(\ref{eq:hyper-systolic_matrix}) is not sufficient to compute the
required forces.  This is because the full data are only stored in the
first row and data pairs constructed using other rows only contain
information about the block particles.

Nevertheless one can construct a different hyper-systolic matrix $\tilde H$ 
that ensures complete force calculations.
For $p=8$ a possible matrix would be:
\begin{equation}
  \label{eq:mat_p=8}
  \tilde H = \left ( \begin{array}{c c c c c c c c}
      \mathbf{1} & \mathbf{2} & \mathbf{3} & \mathbf{4} & 
	\mathbf{5} & \mathbf{6} & \mathbf{7} & \mathbf{8} \\
      8 & 1 & 2 & 3 & 4 & 5 & 6 & 7 \\
      7 & 8 & 1 & 2 & 3 & 4 & 5 & 6 \\
      6 & 7 & 8 & 1 & 2 & 3 & 4 & 5 \\
      5 & 6 & 7 & 8 & 1 & 2 & 3 & 4 \\  
      \mathbf{4} & \mathbf{5} & \mathbf{6} & \mathbf{7} & 
	\mathbf{8} & \mathbf{1} & \mathbf{2} &\mathbf{3} \\ 
    \end{array} \right ) .
\end{equation}
The bold figures represent full data sets and the non-bold figures
indicate the block data sets. With this matrix of shifts, 
one can calculate the complete forces. 
For example, for the first block data set, one uses columns $2, 3, 4$
and $5$. 
The block data can be overwritten as in the systolic algorithm. 
The number of shifts needed for this algorithm is $6$.  First the block
data have to be shifted ($4$ shifts); then they have to be shifted
back to their home processor to add the forces ($1$ shift); and finally
the full data sets have to be updated ($1$ shift).  With the standard
systolic algorithm one would need $8$ shifts. We pay for this
advantage with a higher amount of memory required: one more full data
set has to be saved on each node.

One can construct matrices $\tilde H$ for an arbitrary number of
processors $p$. We call the number of total data sets (bold rows)
$\kappa$ and the number of block data sets $\tilde \kappa$. The number
of rows in the matrix is
\begin{equation}
  \label{eq:kappa_sum}
  \kappa + \tilde \kappa = k.
\end{equation}
To get the full force on each block data set, its contents must be 
compared with every full dataset. 
So each block data set has to meet each full data
set during one of the shifts. 
For each block data set there are $p-1$
full data sets to be met. Thus
\begin{equation}
  \kappa \times \tilde\kappa \geq p-1.
  \label{eq:kappa_combination}
\end{equation}
Equation (\ref{eq:kappa_combination}) allows us to find all possible
combinations of $\kappa$ and $\tilde\kappa$. Table \ref{tab:kappa}
gives the complete list for $p=9$ processing elements.
For all these combinations, one can find a possible hyper-systolic
matrix that guarantees a complete force calculation. 
The $\tilde H$ matrix is
constructed by using the following basis vector:
\begin{equation}
  \label{eq:basisvector}
  A_k = 
  (0, \overbrace{\underbrace{1, 1, \ldots, 1}_{\tilde\kappa 
	\mathrm{\,\,times}}}^{\mathrm{block \, data}}, 
     \overbrace{1, \underbrace{\tilde\kappa, \tilde\kappa, 
	\ldots, 
	\tilde\kappa}_{\kappa-2 \mathrm{\,\,times}}}^{\mathrm{full 
	\, data}}).
\end{equation}
The quantity $\kappa$ determines the amount of memory needed, 
while $k$ determines the amount of communication. 
This is because the number of shifts needed is
$(\tilde\kappa + 1) + (\kappa - 1) = k$ (one shift for each
$\tilde\kappa$-line, one shift for bringing the data to their home
processors and $\kappa-1$ shifts to update the full data sets).

\begin{table}[htbp]
  \begin{center}
    \begin{tabular}{c|c||c}
      $\kappa$ & $\tilde\kappa$ & $k$ \\ \hline
      1 & 8 & 9 \\
      2 & 4 & 6 \\
      3 & 3 & 6 \\
      4 & 2 & 6 \\
      5 & 2 & 7 \\
      6 & 2 & 8 \\
      7 & 2 & 9 \\
      8 & 1 & 9 
    \end{tabular}
    \caption{$\kappa$, $\tilde \kappa$ and $k$ for $p=9$}
    \label{tab:kappa}
  \end{center}
\end{table}

The combination of values in the second line of Table \ref{tab:kappa}
provides the fastest communication. The number of shifts $k$ is
minimal as is the redundancy of data $\kappa$. For optimal memory
usage, the first line is the best choice which leads actually to a
systolic force loop. The last line in Table \ref{tab:kappa} represents
the shared memory approach where a copy of the whole dataset is kept
on each processor. 

We now compute the minimum number of shifts we have to perform
with these algorithms. $\kappa \tilde\kappa$ is approximately $p-1$
(equation \ref{eq:kappa_combination}). Substituting into equation
(\ref{eq:kappa_sum}) we get:
\begin{equation}
  \label{eq:k}
  k = {p-1 \over \kappa} + \kappa.
\end{equation}
The quantity $k$ is the number of shifts that are to be
performed. Thus we want to minimize it:
\begin{equation}
  \label{eq:k_derivative}
  {dk \over d\kappa} = - {p-1 \over \kappa^2}+1 = 0
\end{equation}
and so 
\begin{equation}
  \label{eq:kappa_min}
  \kappa = \sqrt{p-1}.
\end{equation}
Thus we need $\sqrt{p-1}$ full data sets on each processor. 
Substituting (\ref{eq:kappa_min}) into (\ref{eq:k}) we get
\begin{equation}
  \label{eq:k_min}
  k=2\sqrt{p-1}
\end{equation}
and by virtue of (\ref{eq:kappa_sum}),
\begin{equation}
  \label{eq:tildekappa}
  \tilde \kappa = \kappa = \sqrt{p-1}.
\end{equation}

The $\tilde\kappa + 1$ shifts of the block data can be done with the
nonblocking scheme described above.  The $\kappa - 1$ update shifts
have to be done, when the force calculation is completely finished and
therefore cannot be moved to the background.

\subsection{Performance}
\label{sec:performance_hyper-systolic}

The performance of this class of algorithm will now be discussed for
a blocking communication scheme.  The calculation time,
from section \ref{sec:expectation}, is:
\begin{equation}
  \label{eq:hyp_hyp_t_f}
  t_f = {N s \over p} \tau_f.
\end{equation}
The communication time for one shift is $\tau_l + s \tau_c/p$
(equation \ref{eq:t_i_f_c}) assuming that the $s$ block particles are
distributed equally over the processors, so that $s_i = s/p$. The
number of shifts that have to be performed is $k=\kappa+\tilde\kappa$
(equation \ref{eq:kappa_sum}). Thus the total communication time is
\begin{equation}
  \label{eq:hyp_hyp_t_c}
  t_c = k \left ( \tau_l + {s \over p} \tau_c \right ).
\end{equation}
Thus, using a blocking communication scheme, the total time for one
integration step is
\begin{equation}
  \label{eq:hyp_hyp_t}
  t = t_c + t_f 
	\; = \; k \left( 
	\tau_l + {s \over p} \tau_c \right ) 
	+ {N s \over p} \tau_f.
\end{equation}
In the case of minimal communication cost (equation \ref{eq:k_min})
the communication cost is
\begin{equation}
  \label{eq:hyp_hyp_ideal}
  t_c = 2 \sqrt{p} \tau_l + 2 {s \over \sqrt{p}} \tau_c,
\end{equation}
where we approximate $k$ by $2 \sqrt{p}$.  We get the optimal number
of processors that minimize the calculation time by setting the first 
derivative of $p$ with respect to $t$ to zero. This leads us
to
\begin{equation}
  \label{eq:p_optimal}
  p_\mathrm{opt}^{3/2} \tau_l - p_{opt}^{1/2} s \tau_c - N s \tau_f = 0.
\end{equation}
Since the hypersystolic codes are aimed at very massive parallel
machines, we assume that $p$ is a number of order $100$ or
larger. The second term then becomes much smaller than the
first one and: 
\begin{equation}
  \label{eq:optimal_p_again}
  p_\mathrm{opt} \approx \left ( N s {\tau_f \over \tau_l} \right ) ^{2/3}.
\end{equation}
As we measure in our benchmarks, the mean groupsize $s \propto
N^{1/2}$. In this case $p_\mathrm{opt} \approx N (\tau_f /
\tau_l)^{2/3}$. So the optimal processor number is independent of the
communication bandwidth of the host computer while this is not the
case for the systolic algorithm in equation (\ref{eq:p_opt}).

The performance just described is similar to that of the
two-dimensional ring algorithm introdued by Makino \cite{Makino:2001}.
If we choose $k=2 \sqrt{p}$ we get the same amount of communication
time and the same amount of communication as his algorithm needs.
Makino's algorithm however is not strictly hyper-systolic.  It is
designed for a 2D network of processors and the systolic shifts are
carried out within rows and columns of this 2D array.  Such an
algorithm is far less flexible than the hyper-systolic algorithm
described above, for the follwing reasons.  First, it only works for
processor numbers $p$ for which $\sqrt{p}$ is an integer.
Second, Makino's algorithm corresponds to our hyper-systolic algorithm
with $\kappa = \tilde\kappa = \sqrt{p}$. Therefore it requires that
$\sqrt{p}$ full data arrays are saved on each processor. Our algorithm
works with less memory without losing communication efficiency. For
example, for $p=9$, Makino's algorithm needs to save $3$ datasets on
each processor, while with the hyper-systolic algorithm, we can choose
$\kappa = 2$ and $\tilde\kappa = 4$ (Table
\ref{tab:kappa}). $k$ still equals $6$. Thus the communication
complexity is the same as in the case $\kappa=\tilde\kappa=3$
while the memory requirements are reduced by one. Also the number of shifts
that cannot be done in a non-blocking way ($\kappa-1$) is reduced.

\subsection{Small group sizes and the hyper-systolic algorithm}
\label{sec:small_group_sizes_and_the_hyper-systolic_algorithm}
Above we discussed the effect of small group sizes on systolic algorithms. 
We defined a group to be small if not all processors provide particles to the
group.  We found that the systolic algorithm with nonblocking communication is
able to deal with small group sizes in such a way that, 
over a large percentage of time, the force calculation is running parallel.

In a hyper-systolic scheme, the shifts of the block data can be done
with blocking or nonblocking routines. Thus the same considerations as
in section \ref{sec:the_effect_of_small_group_sizes} apply. In
addition we encounter a new problem: In the hyper-systolic algorithm
the particle data are not sent to each processor. For example in the
case of $p=8$ and a hyper-systolic matrix like in equation
(\ref{eq:mat_p=8}), calculations for particles provided by processor
$1$ are only done on processors $2, 3, 4$ and $5$. If we only have one
particle in a group, and this is provided by processor $1$, then only
these four processors do calculations, while the remaining four
processors wait until the next integration step. The performance would
be comparable to a $4$ processor parallel computation.

In general, if one chooses a hyper-systolic algorithm with the
smallest communication complexity $\tilde \kappa = \sqrt{p-1}$
(equation \ref{eq:tildekappa}) and a group size of one, then only
$\sqrt{p-1}$ processors are doing the computation.

If more than one but not all processors contribute particles to the
group, the parallelization depends strongly on the specific processor
number. For example, in the case of matrix \ref{eq:mat_p=8}, if
processors $1$ and $2$ provide particles, $5$ processors are working
in parallel. If processors $1$ and $5$ provide particles, then all of the
$8$ processors are working in parallel.

\section{An Application: Gravitational Brownian Motion}
\label{sec:brown}

\subsection{The Problem}

The algorithms described here are ideally suited to problems
requiring large $N$ and small to moderate numbers of integration steps.  
One such problem is the Brownian motion of a massive point particle 
that responds to the gravitational force from $N$, less-massive
particles.
Let $\mh$ and ${\bf R}$ be the mass and position of the Brownian particle, 
$m$ the mass of a perturber, and ${\bf r}_j$ the position of the $j$th
perturber particle.
The equations of motion are
\begin{eqnarray}
{\bf \ddot R} &=& -Gm\sum_{k=1}^N{\left({\bf R}-{\bf r}_k\right)\over 
|{\bf R}-{\bf r}_k|^3},\label{eq:motion1} \\
{\bf \ddot r}_j &=& -Gm\sum_{k=1}^N{\left({\bf r}_j-{\bf r}_k\right)\over 
|{\bf r}_j-{\bf r}_k|^3}  + G\mh{\left({\bf R}-{\bf r}_j\right)\over 
|{\bf R}-{\bf r}_j|^3},
\label{eq:motion2}
\end{eqnarray}
where the summation in equation (\ref{eq:motion2}) excludes $k=j$.
The total mass of the stellar system excluding the Brownian particle
is $Nm\equiv M$.

This problem is relevant to the behavior of supermassive
black holes at the centers of galaxies. 
Black holes have masses of order $10^6M_{\odot}\lap \mh\lap 10^9M_{\odot}$, 
with $M_{\odot}$ the mass of the sun, 
compared with a total mass of a galactic nucleus of
$\sim 10^9M_{\odot}$ and of an entire galaxy, $\sim 10^{11}M_{\odot}$.
By analogy with the fluid case, 
the rms Brownian velocity of the black hole is expected to be of order
\begin{equation}
V_{rms}\approx\sqrt{3}\sqrt{m\over \mh}\sigma
\label{eq:brown1}
\end{equation}
where $\sigma$ is the 1D velocity dispersion of stars in the vicinity
of the black hole and $m$ is a typical stellar mass.
But Brownian motion in a self-gravitating system is expected to
differ from that in a fluid, for a variety of reasons.
A massive particle alters the potential when it moves and may
excite collective modes in the background.
Its effective mass may differ from $\mh$ due to particles bound to it,
and the force perturbations acting on it are not necessarily localized
in time.
For these reasons, it is important to treat the motion of the background
particles in a fully self-consistent way, as in equations 
(\ref{eq:motion1}-\ref{eq:motion2}).

Integrations involving large particle numbers are slow
even with a parallel agorithm,
but fortunately the time required for a massive particle 
(the ``black hole'') to reach
a state of energy equipartition with the stars, $T_{eq}$, is expected
to be less than a single crossing time of the stellar system in which 
it is imbedded.
We derive this result as follows. 
Starting from rest, the mean square velocity of the black hole
should evolve approximately as
\begin{eqnarray}
\langle V^2\rangle \approx \dvp t
\end{eqnarray}
where
\begin{equation}
\dvp = {8\sqrt{2\pi}\over 3}{G^2m\rho\ln\Lambda\over\sigma}
\end{equation}
(e.g. \cite{Brownian:01}).
Here $\rho$ is the mass density of stars and
$\ln\Lambda$ is the Coulomb logarithm, which is of order
unity for this case (\cite{Nuclei:01}) and will henceforth be
set equal to one.
The time $T_{eq}$ required for $\langle V^2\rangle$ to 
reach its expected equilibrium value of $\sim(m/M)\sigma^2$
is therefore
\begin{equation}
T_{eq} \approx {m\over \mh}{\sigma^2\over\dvp}
\approx {\sigma^3\over G^2\mh\rho}.
\end{equation}
This may be written
\begin{equation}
T_{eq}\approx T_D\left({M_{gal}\over \mh}\right)\left({\rho\over\langle\rho\rangle}\right)^{-1}
\end{equation}
where $\langle\rho\rangle$ is the mean density of the galaxy within its
half-mass radius $R_{1/2}$ and $T_D\equiv R_{1/2}/\sigma$, the crossing
time within that radius.

\begin{figure}
\includegraphics[width=\linewidth]{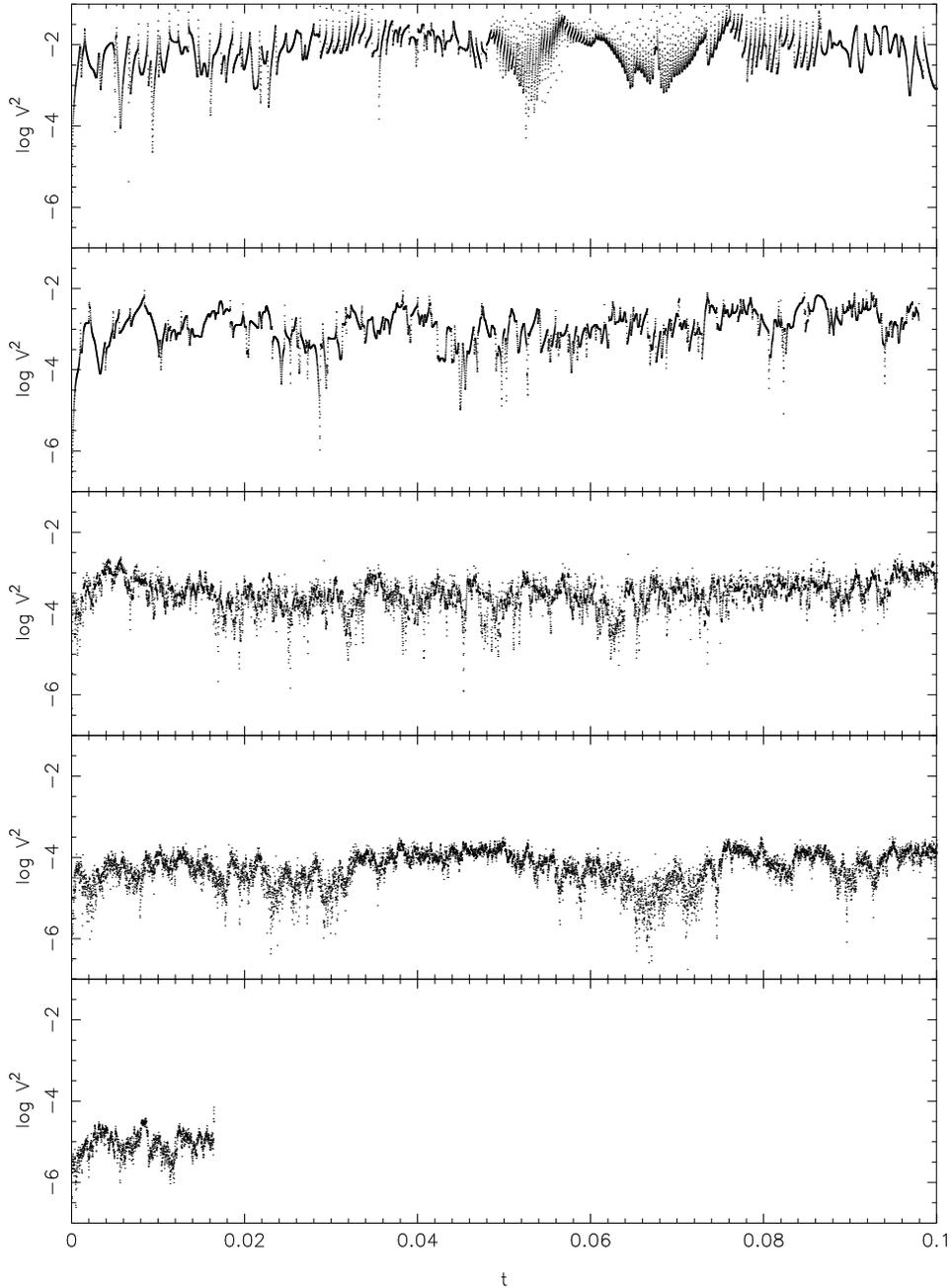}
\caption{
Time evolution of the squared velocity of the massive (``black hole'')
particle in the Brownian motion experiments.
The background stellar system consists of $N=10^6$ particles
of mass $m=10^{-6}$ distributed according to a Dehnen density law,
Equation (\ref{eq:dehnen2}).
The mass $\mh$ of the black hole particle increases
downward by a factor of 10 between each frame, from $\mh=10^{-5}$ at the
top to $\mh=10^{-1}$ at the bottom.
The black hole particle is started at rest and rapidly comes into energy
equipartition with the lighter particles.
The amplitude of its random motion varies inversely with its mass
as expected from energy equipartition arguments.
A close binary was formed after 0.016 time units in the simulation with
the most massive black hole particle, requiring extremely small
timesteps and forcing the run to stop.
}
\label{fig:evolve}
\end{figure}

A typical environment for a supermassive black hole is at the center of
a galaxy in which the stellar density varies as
\begin{equation}
\rho \approx \langle\rho\rangle\left({r\over R_{1/2}}\right)^{-\gamma}
\end{equation}
with $1\lap\gamma\lap 2$ \cite{Gebhardt:96}.
Thus
\begin{equation}
T_{eq}\approx T_D\left({M_{gal}\over \mh}\right)\left({r\over R_{1/2}}\right)^{\gamma}.
\end{equation}
To estimate the second factor in parentheses, 
we assume that the stars that most strongly
affect the motion of the black hole are within a distance
$r\approx G\mh/\sigma^2\approx (\mh/M_{gal})R_{1/2}$, the classical
radius of influence of the black hole \cite{Peebles:72}.
Then
\begin{equation}
T_{eq}\approx T_D\left({\mh\over M_{gal}}\right)^{0.5}
\end{equation}
where $\gamma$ has been set equal to $1.5$, roughly the value 
characterizing the stellar density cusp surrounding the Milky Way black 
hole \cite{Alexander:99}.
Setting $\mh/M_{gal}\sim 10^{-3}$ \cite{Merritt:01}, we find $T_{eq}\sim T_D/30$.
While this result is very approximate, it suggests that a massive
particle will reach energy equipartition with the ``stars'' in much
less than one crossing time of the stellar system.
Note that this result is independent of the masses of the perturbing
particles and hence of $N$. 
However the predicted value of the equilibrium velocity dispersion 
of the black hole does depend on $m/\mh$ (cf. equation \ref{eq:brown1}).
We evaluate this dependence below via numerical experiments.

\subsection{The Experiments}

We constructed initial conditions by distributing $N=10^6$, equal-mass
particles representing the stars according to Dehnen's law 
(\ref{eq:dehnen}) with $\gamma=1.5$,
\begin{equation}
\rho(r) = \frac{3 M}{8 \pi} \; \frac{a}{r^{1.5} \left(r + a\right)^{2.5}}
\label{eq:dehnen2}
\end{equation}
and a central point of mass $\mh$ representing the black hole.
The value chosen for the slope of the central density cusp,
$\gamma=1.5$, is similar to the value near the center of the Milky
Way galaxy \cite{Alexander:99}.
The velocities of the ``star'' particles were selected from the
unique isotropic distribution function that reproduces a steady-state
Dehnen density law in the presence of a central point mass
\cite{Tremaine:94}.
The initial velocity of the black hole particle was set to zero.
We carried out integrations for five different values of the black
hole mass, $\mh=(0.00001, 0.0001, 0.001, 0.01, 0.1)$ in units where the
total mass $M$ in ``stars'' is unity. 
The gravitational constant $G$ and the Dehnen-model scale length $a$
were also set to one.
Integrations were carried out until a time of $t_\mathrm{max} = 0.1$, 
compared with a crossing time for the overall system of $a/\sigma\approx 1$.
In the case of the largest black hole mass tested by us, 
$\mh=0.1$, the high velocity of
stars in the density cusp around the massive particle required the use
of very small time steps for many of the particles.
This run was not continued beyond $t\approx 0.016$.
All runs were carried out using 128 processors on the Cray T3E 900
supercomputer in Stuttgart.

\begin{figure}
\includegraphics[width=\linewidth]{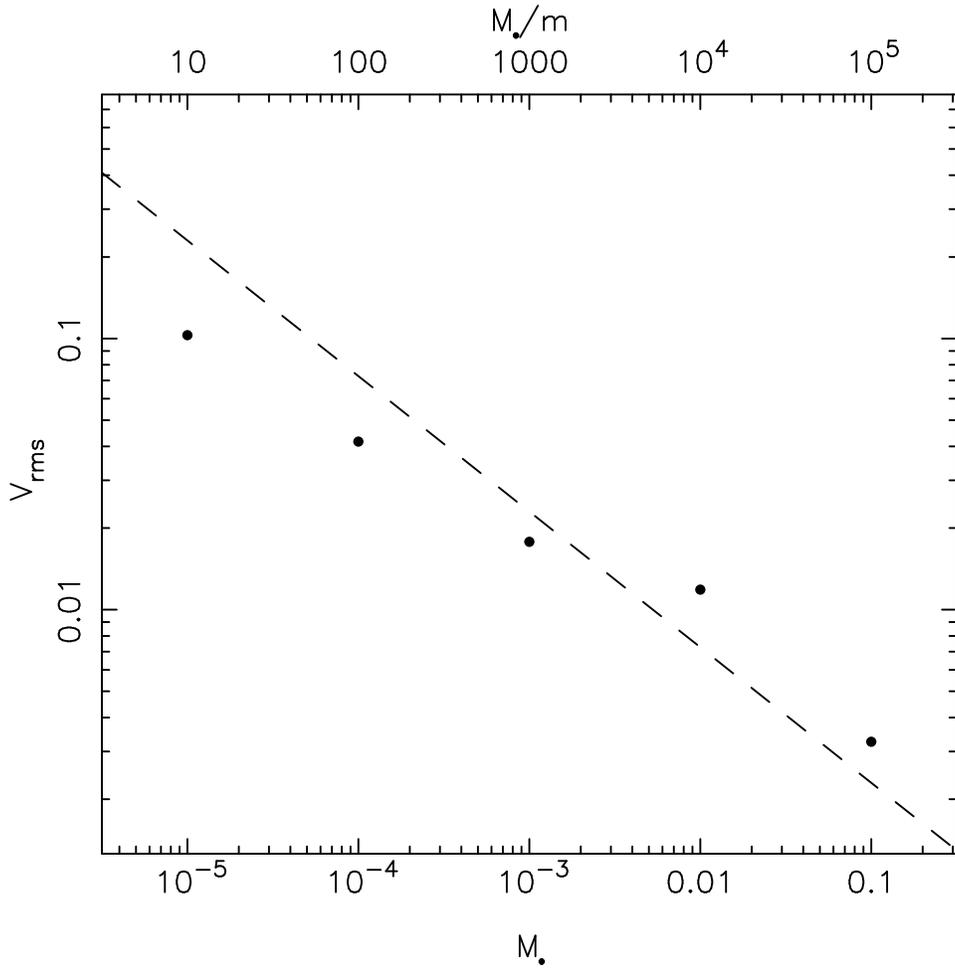}
\caption{
RMS velocity of the ``black hole'' particle in the Brownian motion experiments.
$V_{rms}$ was computed using the average value of $V^2(t)$ between $t=0$
and $t=0.1$, with the exception of the point at $\mh=0.1$ for which
the integration time was shorter (cf. Figure \ref{fig:evolve}).
Dashed line is the predicted relation based on energy equipartition arguments
(see text).
$m$ is the mass of a ``star'' particle.
}
\label{fig:equipart}
\end{figure}

Figure \ref{fig:evolve} shows the time dependence of the squared
velocity of the black hole particle, $V^2(t)$, for each of the runs.
The motion appears to reach a statistical steady state in much
less than one crossing time, as expected,
and the inverse dependence of $\langle V^2\rangle$ on $\mh$ is apparent.
There were no discernable changes in the spatial or velocity
distribution of the ``star'' particles during these runs
outside of the region where the black hole wandered.

Figure \ref{fig:equipart} shows $V_{rms}$ as a function of $\mh$ for each 
of the runs.
$V_{rms}$ was computed by averaging $V^2(t)$ over the full integration
interval.
From equation (\ref{eq:brown1}), and setting $m=1/N=10^{-6}$, we expect
\begin{equation}
V_{rms}\approx 1.73\times 10^{-3} \sigma \mh^{-0.5}
\label{eq:vrms}
\end{equation}
where $\sigma$ is the 1D velocity dispersion of stars in the vicinity
of the black hole particle.
However it is not clear what value of $\sigma$ should be used 
in this formula, since the velocity dispersion of the stars
is a function of radius in the initial models:
$\sigma$ increases toward the center to a maximum of $\sim 0.42$ at
a radius of $\sim 0.2$, then drops slowly inward before rising
again as $\sigma^2 \approx 2/5r$ within the radius of influence
of the black hole.
Using the peak value of $\sigma$ in the black-hole-free Dehnen model,
$\sigma=0.42$, equation (\ref{eq:vrms}) predicts
\begin{equation}
V_{rms}\approx 7.27\times 10^{-4} {\mh}^{-0.5}.
\end{equation}
This line is shown in Figure \ref{fig:equipart}.
The measured values of $V_{rms}$ are consistent with this
prediction, although there is a hint that the dependence of $V_{rms}$ on
$\mh$ is slightly less steep than $\mh^{-0.5}$. 
This result suggests that gravitational Brownian motion
may be very similar to its fluid counterpart.
In particular, we note that $V_{rms}$ is correctly
predicted by equation (\ref{eq:brown1}) if one uses a value for
$\sigma$ in that equation that is measured well outside of the 
region of gravitational influence of the black hole.
This suggests that most of the perturbations leading to the
black hole's motion come from distant stars.

We can use our results to estimate the random velocity of the
supermassive black hole at the center of the Milky Way galaxy.
It has recently become feasible to measure the motion of the Milky
Way black hole \cite{Backer:99,Reid:99}, whose mass
is $\mh\approx 2.5-3.5\times 10^6M_{\odot}$
\cite{Ghez:98,Genzel:00,Chakrabarty:01}, 
roughly $10^{-3}$ times the mass of the
Milky Way bulge \cite{Merritt:01}.
The masses of stars in the cluster surrounding the black hole
are of order $10-20\msun$ \cite{Genzel:97}.
We adopt $\sigma\approx 100$ km s$^{-1}$, 
roughly the peak velocity dispersion measured for the
stars in the Milky Way bulge outside of the region of influence
of the black hole \cite{Kent:92}; here we make use of our result that
the Brownian velocity in the $N$-body simulations is correctly
predicted by equation (\ref{eq:brown1}) when $\sigma$ is replaced by
its peak value outside of the region of influence of the central black hole.
The resulting prediction for the 3D random velocity of the Milky Way 
black hole is
\begin{equation}
V_{rms}\approx 0.40 \left({m_*\over 15\msun}\right)^{1/2} \left({\mh\over 3\times 10^6\msun}\right)^{-1/2} \mathrm{km}~\mathrm{s}^{-1}.
\end{equation}
The predicted velocity could be much greater than $0.4$ km s$^{-1}$
if the objects providing the force perturbations are much more
massive than $15\msun$, e.g., giant molecular clouds.
Current upper limits on the proper motion (2D) velocity of the Milky Way black
hole are $\sim 20$ km s$^{-1}$ \cite{Backer:99,Reid:99}.

\section{Conclusions}

We have introduced two variants of a systolic algorithm for parallelizing
direct-summation $N$-body codes implementing individual block time step 
integrators: the first with blocking communication, and the second with
non-blocking communication.
Performance tests were carried out using $N$-body models similar
to those commonly studied by dynamical astronomers, in which
the gravitational forces vary widely between core and halo and for
which the particle block sizes are typically very small.
The nonblocking scheme was found to provide far better performance 
than the blocking scheme for such systems,
providing a nearly ideal speedup for the force calculations. 
By engaging a sufficient number of computing nodes, particle numbers
in excess of $10^6$ are now feasible for direct $N$-body simulations.
For parallel machines with very large processor numbers, we describe the
implementation of a hyper-systolic computing scheme which provides a
communication scaling of $O(\sqrt{p})$ at the expense of increased
memory demands.

The codes used to write this paper are available for download at:\\
\verb+http://www.physics.rutgers.edu/~marchems/download.html+ 

\begin{acknowledgment}

This work was supported by NSF grant 00-71099 and by
NASA grants NAG5-6037 and NAG5-9046 to DM.
We thank Th.~Lippert, W.~Schroers, and K.~Schilling for
their encouragement and help. 
We are grateful to the NASA Center for Computational Sciences
at NASA-Goddard Space Flight Center, 
the National Computational Science Alliance,
the Center for Advanced Information Processing at Rutgers University,
the John von Neumman Institut f\"ur Computing in J\"ulich, and the 
 H\"ochstleistungsrechenzentrum in Stuttgart
 for their generous allocations of computer time.

\end{acknowledgment}

\end{article}
\end{document}